\documentclass[twocolumn,prd]{revtex4}
\usepackage{graphicx}
\usepackage{latexsym}
\usepackage{amsmath}

\newcommand{\beq}{\begin{equation}}
\newcommand{\eeq}{\end{equation}}
\newcommand{\bea}{\begin{eqnarray}}
\newcommand{\eea}{\end{eqnarray}}

\begin{document}

\title{On the $s-\bar{s}$ asymmetry in the nucleon sea}

\date{\today}

\author{C. Avila}
\author{I. Monroy}
\author{J.C. Sanabria}
\email{jsanabri@uniandes.edu.co}
\affiliation{Departamento de F\'{\i}sica, Universidad de los Andes, A.A.
4976, Bogota, D.C., Colombia}

\author{J. Magnin}
\email{jmagnin@cbpf.br}
\affiliation{Centro Brasileiro de Pesquisas F\'{\i}sicas, Rua Dr. Xavier
Sigaud 150, Urca 22290-180, Rio de Janeiro, Brazil}

\begin{abstract}
We study the $s-\bar s$ asymmetry in the nucleon sea using a model in
which the proton wave function includes a Kaon-Hyperon Fock
state. Parameters of the model are fixed by fitting the $s-\bar s$
asymmetry obtained from global fits to Deep Inelastic Scattering data. 
We discuss possible effects of the $s-\bar s$ asymmetry on the measurement
of the Weinberg angle by the NuTeV Collaboration.
\end{abstract}

\maketitle

\section{Introduction}

The first studies about a possible asymmetry in the strange sea of
the nucleon dates from 1987, when Signal and Thomas~\cite{signal}
discused the possibility of a $K^+\Lambda$ pair component in the
proton wavefunction. Since then on, several models have been propossed
for the nucleon structure, allowing for an asymmetric  $s-\bar s$
sea~\cite{varios,mane,christiansen}. However, no experimental evidence 
was presented on this subject until the global fit of Deep Inelastic Scattering 
(DIS) data by Barone {\it et al.}~\cite{barone} in 2000. Most recently, the
$s-\bar s$ asymmetry in the nucleon sea was called for as a possible
explanation~\cite{explanations} for the almost $3~\sigma$ difference between the NuTeV
$sin^2\theta_W$ result~\cite{zeller} and global fits~\cite{LEP}.

From a theoretical point of view, it is interesting to note that although sea quarks in
the nucleon originating in gluon splitting necessarily have symmetric
momentum distributions, after interacting with the valence quarks and
the remaining partons in the sea, their momentum distributions do not
have to be equal. This can be interpreted as the formation of a
virtual $K^+\Lambda$ pair in the nucleon structure. Being this the
case, it is easy to see that, since the $s$ and the $\bar s$ quarks
are part of the $\Lambda$ and $K^+$ respectively, then their momentum
distributions would be different. This difference, which is merely a
consequence of the interaction of sea quarks with the remaining partons
in the nucleon, has to be understood as part of the
non-perturbative dynamics responsible for the formation of the nucleon
as a bound state of quarks
and gluons. Recall also the ${\bar u}-{\bar d}$ asymmetry and the
Gottfried Sum Rule violation, known since the 
New Muon Collaboration results~\cite{nmc}, which can also been
explained in terms of a $n\pi^+$ and a $\Delta^{++}\pi^-$ components
in the proton wave function~\cite{magnin}. It is interesting to note 
however, that a small $s-\bar s$ asymmetry arises as a NNLO
perturbative effect~\cite{deflorian}, nevertheless, the integrated
value, $S=\int_0^1{x[s(x)-{\bar s}(x)]dx}\simeq -5\times10^{-4}$ at $Q^2=20$ GeV$^2$, is
too small and negative to account for the NuTeV result.

In this work, we shall consider a model for the extrange sea of the
proton which can describe the form of the $s-\bar s$ asymmetry
extracted from global fits to DIS data. After fixing the parameters of
the model, in Section~\ref{sec3}, we shall study the effect of this
asymmetry, together with possible effects coming from the non
isoscalarity of the target and nuclear medium effects, in the determination 
of $\sin ^2\theta_W$ by the NuTeV Collaboration. Section~\ref{conclu} will be 
devoted to discussion and conclusions.

\section{A model for the $s(x)-\bar{s}(x)$ asymmetry}

Different models~\cite{signal,varios,mane,christiansen} have attempted to
predict the $s-\bar s$ asymmetry.
Among them, the most promising approach seems to be the Meson Cloud
Model (MCM).  In the MCM fluctuations of the proton to kaon-hyperon
virtual states are responsible for the $s-\bar s$ asymmetry. Since the $s$
quark belong to the hyperon and the $\bar s$ quark to the kaon, the
asymmetry arises naturally due to the different momentum
carried by the kaon and the hyperon in the fluctuation. 

Two different approaches exist within the MCM. The
first is based in a description of the form factor of the extended 
proton-kaon-hyperon vertex~\cite{signal,mane} and, the second one, in terms
of parton degrees of freedom~\cite{christiansen}. In the first
approach, the knowledge of the  form
factors is crucial to get a reasonable description of the $s-\bar s$
asymmetry (see e.g. Ref.~\cite{mane}). In the
second one, fluctuations are generated through gluon emission from 
the constituent valence quarks and its subsequent splitting to a 
$s-\bar s$ pair~\cite{christiansen}. This $s-\bar s$ pair then
recombines with constituent quarks to form a kaon-hyperon bound
state. In what follows, we will adopt the second approach.

\subsection{The model}

We start by considering a simple picture of the nucleon in the 
infinite momentum frame as being formed by three dressed valence 
quarks - {\it valons}, $v(x)$ - which carry all of its momentum~\cite{hwa}. 

In the framework of the MCM, the nucleon can fluctuate to a  
meson-baryon bound state carrying zero net strangeness. As a first step
in such a process, we may consider that  
each valon can emit a gluon which, before interacting, 
decays perturbatively into a $s\bar s$ pair. The probability of 
having such a perturbative $q\bar{q}$ pair can then be computed in terms 
of Altarelli-Parisi splitting functions~\cite{alta-par} 
\bea
 P_{gq} (z) &=& \frac{4}{3} \frac{1+(1-z)^2}{z}, \nonumber \\
P_{qg} (z) &=& \frac{1}{2} \left( z^2 + (1-z)^2 \right).
\label{eq2}
\eea
These functions have a physical interpretation as the probability 
of gluon emision and $q\bar{q}$ creation with momentum fraction $z$ 
from a parent quark or gluon respectively. Hence, 
\bea
q(x,Q^2) =&& \bar{q}(x,Q^2) = N \frac{\alpha_{st}^2(Q^2)}{(2\pi)^2}
\times \nonumber \\
&&\int_x^1 {\frac{dy}{y} P_{qg}\left(\frac{x}{y}\right) 
\int_y^1{\frac{dz}{z} P_{gq}\left(\frac{y}{z}\right) v(z)}}
\label{eq3}
\eea
is the joint probability density of obtaining a quark or anti-quark 
coming from  subsequent decays $v \rightarrow v + g$ 
and $g \rightarrow q + \bar{q}$ at some fixed low $Q^2$. 
As the valon distribution does not depend on $Q^2$ \cite{hwa}, 
the scale dependence in eq.~(\ref{eq3}) only exhibits through the 
strong coupling constant $\alpha_{st}$. The range of values of $Q^2$ 
at which the process of virtual pair creation occurs in this approach
is typically below 1 GeV$^2$, as dictated by 
the valon model of the nucleon. For definiteness, 
we will use $Q = 0.7$ GeV as in Ref.~\cite{hwa}, for which 
$\alpha_{st}^2 \sim 0.3$ is still sufficiently small to allow for 
a perturbative evaluation of the $q\bar{q}$ pair production.
Since the scale must be consistent with the valon picture,
the value of $Q^2$ is not really free and cannot be used
to control the flavor produced at the $gq\bar{q}$ 
vertex. Instead, this role can be ascribed to the normalization constant
$N$, which must be such that to a heavier quark corresponds a lower 
value of $N$.

Once a $s\bar{s}$ pair is produced, it can rearrange itself with the 
remaining valons so as to form a most energetically favored 
meson-baryon bound state. When the nucleon fluctuates 
into a meson-baryon bound state, the meson and baryon probability
densities inside the nucleon are not independent. Actually, to ensure the zero 
net strangeness of the nucleon and momentum conservation, the in-nucleon
meson and baryon distributions must fulfill two basic constraints,
\bea
\label{eq4a}
\int_0^1 {dx \left[P_B(x) - P_M(x) \right]} &=&  0, \\
\int_0^1 {dx \left[xP_B(x) + xP_M(x) \right]}  &=&  1,
\label{eq4}
\eea
for all momentum fractions $x$. 

The meson, $P_M(x)$, and baryon,
$P_B(x)$, probability density functions have to be calculated by means
of effective techniques in order to deal with the non-perturbative QCD
processes inherent to the dressing of quarks into hadrons. In
Ref.~\cite{christiansen}, these probability densities have been
related to the cross section for meson production by recombination, and
the model by Das and Hwa~\cite{das-hwa} was used to obtain them. In
this work, since the aim is to compare the model to
experimental data by means of a fit, we will follow a different
approach. Let us note that, for reasonable valon distributions in the nucleon and 
sea quark distributions of the form given by eq.~(\ref{eq3}), 
the result of using the recombination model gives for the meson
probability density a function of the form $x^a(1-x)^b$. Then we will
assume
\beq
P_M(x) = \frac{1}{\beta(a_{\mbox{\tiny KN}}+1,b_{\mbox{\tiny KN}+1)}} x^{a_{\mbox{\tiny
    KN}}}(1-x)^{b_{\mbox{\tiny KN}}}, 
\label{pmpdf}
\eeq
which is properly normalized to one. If for the in-nucleon baryon
probability we use the same functional form as for the meson, 
\beq
P_B(x) = \frac{1}{\beta(a_{\mbox{\tiny HN}}+1,b_{\mbox{\tiny HN}}+1)} x^{a_{\mbox{\tiny
    HN}}}(1-x)^{b_{\mbox{\tiny HN}}}, 
\label{pbpdf}
\eeq
it is automatically satisfied the requeriment of zero net
strangenes. In addition, interpreting $a_{\mbox{\tiny KN}},~b_{\mbox{\tiny KN}},
~a_{\mbox{\tiny HN}}$ and $b_{\mbox{\tiny HN}}$ as parameters of the model, and
recognizing that eq.~(\ref{eq4}) fix one of them as a function of the
remaining three by means of
\bea
\label{sum}
&&\frac{\Gamma(a_{\mbox{\tiny KN}}+b_{\mbox{\tiny KN}}+2)\Gamma(a_{\mbox{\tiny KN}}+2)}
{\Gamma(a_{\mbox{\tiny KN}}+1)\Gamma(a_{\mbox{\tiny KN}}+b_{\mbox{\tiny
      KN}}+3)} + \nonumber \\
&&\frac{\Gamma(a_{\mbox{\tiny HN}}+b_{\mbox{\tiny HN}}+2)\Gamma(a_{\mbox{\tiny HN}}+2)}
{\Gamma(a_{\mbox{\tiny HN}}+1)\Gamma(a_{\mbox{\tiny HN}}+b_{\mbox{\tiny
      HN}}+3)} = 1\;,
\eea
then the momentum conservation sum rule is also fulfilled.

The non-perturbative strange and anti-strange sea distributions in the
nucleon can be now computed by means of the two-level convolution formulas  
\bea
\label{eq8a}
s^{NP}(x) &=& \int^1_x {\frac{dy}{y} P_B(y)\ s_{H}(x/y)}  \\
\bar{s}^{NP}(x) &=& \int^1_x {\frac{dy}{y} P_M(y)\ \bar{s}_{K}(x/y)},
\label{eq8}
\eea
where the sources $s_{B}(x)$ and $\bar{s}_{M}(x)$ are primarily the 
probability densities of the strange valence quark and anti-quark 
in the baryon and meson respectively, evaluated at the hadronic 
scale $Q^2$ \cite{signal}.
In principle, to obtain the non-perturbative distributions given by 
eqs.~(\ref{eq8}), one should sum over all the strange meson-baryon 
fluctuations of the nucleon but, since such hadronic Fock states are 
necessarilly off-shell, the most likely configurations are 
those closest to the nucleon energy-shell, namely 
$\Lambda^0K^+$, $\Sigma^+K^0$ and $\Sigma^0K^+$, for a 
proton state.

\subsection{Fit to $xs(x)-x\bar{s}(x)$ data}

In order to fit to experimental data on the $s-\bar s$ asymmetry and to
extract the parameters of the model, we will use
\bea
\bar{s}_K(x) &=& \frac{1}{\beta(a_{\mbox{\tiny K}}+1,b_{\mbox{\tiny
	K}}+1)}x^{a_{\mbox{\tiny K}}}(1-x)^{b_{\mbox{\tiny K}}}\;,\\
s_H(x) &=& \frac{1}{\beta(a_{\mbox{\tiny H}}+1,b_{\mbox{\tiny
	H}}+1)}x^{a_{\mbox{\tiny H}}}(1-x)^{b_{\mbox{\tiny H}}}\; ,
\eea
which are consistent with the hypothesis that the in-meson and baryon
are formed by valons. Then the $s-\bar s$ asymmetry of the nucleon is
given by
\beq
xs(x)-x\bar{s}(x) = N^2\left[xs^{NP}(x)-x\bar{s}^{NP}(x)\right]\;,
\label{eq11}
\eeq
at the valon scale $Q^2=0.49$ GeV$^2$, and where $N^2$ is the
probability of the $\left|KH\right>$ Fock state in the proton wave
function, which is related to the probability of having a $s-\bar s$
quark pair out of eq.~(\ref{eq3}). 

The model has then a total of 8 parameters to be fixed by
fits to experimental data. 

\begin{figure}[t]
\includegraphics[scale=0.45]{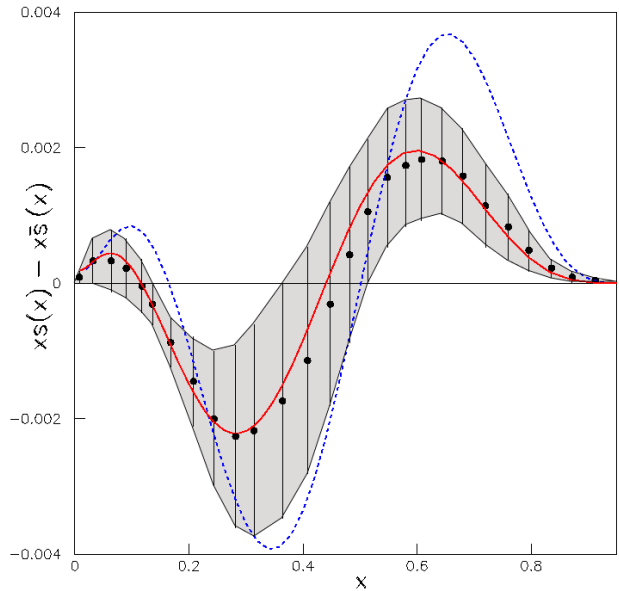}
\caption{\label{fig1}The model compared to experimental data at
  $Q^2=20$ GeV$^2$ (full line). The
  curve is the result of the fit, data points were extracted to fit in the 
  shadowed region as given in Ref.~\cite{portheault} (see the text). The dashed 
  line is the model at $Q^2=0.49$ GeV$^2$.}
\end{figure}

Results of the fit to experimental data from Ref.~\cite{portheault} are
shown in Fig.~\ref{fig1} and Table~\ref{table1}. For the experimental
data, we extracted 27 segments from the allowed band in Figure 2 of 
Ref.~\cite{portheault} and assumed that the midpoint of each segment
is the most probable value which we interpreted as the value for the
asymmetry, while the half lenght of the segment was interpreted as the error
bar. Notice however that this procedure has to be taken only as a way to fit 
our model inside the allowed region for the $s-\bar{s}$ asymmetry.

The fit was performed by minimizing the $\chi^2$
using MINUIT. In the fitting procedure, since the allowed bars for the
$s-\bar s$ asymmetry are given at $Q^2=20$ GeV$^2$, the parameters
where chosen, then the asymmetry was evolved from $Q^2=0.49$ GeV$^2$
to $Q^2=20$ GeV$^2$, the $\chi^2$ was evaluated and the procedure was
repeated until a minimum was reached. 

The evolution has been done at
NLO and a value of $S^-  = 0.87\times 10^{-4}$ at 
$Q^2= 20$ GeV$^2$ has been obtained, which is also the average $Q^2$ reported by
NuTeV~\cite{zeller,zeller2}. Although this value of $S^-$ is small to
account for the anomalous result for $\sin^2\theta_W$ reported by
NuTeV, it is positive. It is also conceivable that, by performing a
NNLO evolution, the negative contribution of the perturbative
asymmetry in $s-\bar s$ be compensated by a bigger positive
non-perturbative asymmetry. 

\begin{table}[t]
\begin{ruledtabular}
\begin{tabular}{cc}
Parameter & Value \\
\hline
$a_{KN}$    & 2.06 $\pm$  2.62$\times 10^{-7}$ \\
$b_{KN}$    & 2.14 $\pm$  0.11 \\
$a_K$       & 5.14 $\pm$  1.93 \\
$b_K$       & 0.90 $\pm$  0.34 \\
$a_{HN}$    & 1.17 $\pm$  0.35 \\
$a_H$       & 9.47 $\pm$  0.61 \\
$b_H$       & 2.51 $\pm$  0.61 \\
$N^2$       & 0.04 $\pm$  0.02 \\
\end{tabular}
\end{ruledtabular}
\caption{\label{table1}Fit results. Parameters $a_{KN}, b_{KN}$ are
  for the Kaon probability density in the nucleon, $a_{HN}, b_{HN}$
  for the Hyperon probability density, $a_K, b_K$ for the anti-strange
  valon probability density in the Kaon and $a_H, b_H$ for the strange
  valon density in the Hyperon. The $\chi^2$/d.o.f. = 0.15 and $b_{HN}$
  = 1.11 is the result of the constraint due to the momentum sum rule
  of eq.~(\ref{sum})}
\end{table}

\begin{figure}[t]
\includegraphics[scale=0.45]{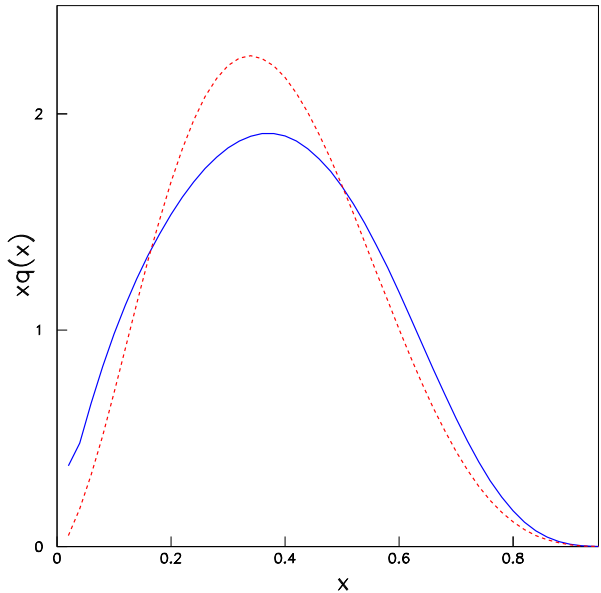}
\caption{\label{fig2}The $xs(x)$ (full line) and $x\bar{s}(x)$ (dashed line) at $Q^2=0.49$ 
  GeV$^2$ as given by the model.}
\end{figure}

In Fig.~\ref{fig2}, the $xs$ and $x\bar{s}$ distributions are displayed at the $Q^2$ 
scale where the evolution starts, namely, $Q^2=0.49$ GeV$^2$.
\newline

\section{The effect of $s(x)-\bar s(x)$ on the determination of
  $\sin^2\theta_W$}
\label{sec3}

The weak mixing angle is one of the basic parameters of the
standard model of electroweak interactions.  An experimental value
for $\sin^{2}\theta_{W}$ has been obtained, using the
\emph{on-shell} renormalization scheme, from a global fit to the
precise electroweak measurements performed by the LEP experiments
at CERN and the SLD experiment at SLC, together with data from
several other experiments at Fermilab\cite{LEP}. This global
analysis lends a value of
\begin{equation*}
\sin^{2}\theta_{W} = 0.2227 \pm 0.0004\,,
\label{global}
\end{equation*}
excluding the data from CCFR and NuTeV experiments.

The NuTeV collaboration reported a value of $\sin^{2}\theta_{W}$
extracted from the analysis of neutrino and antineutrino charged
current (CC) and neutral current (NC) scattering data\cite{zeller}.
The \emph{on shell} value obtained by this experiment is
\begin{equation*}
\sin^{2}\theta_{W} = 0.22773 \pm 0.00135\text{(stat)} \pm
0.00093\text{(syst)}\,;
\end{equation*}
which is $\sim 3\sigma$ away from the the electroweak global fit
value.

Due to an inevitable contamination of the
$\nu_{\mu}$($\bar{\nu}_{\mu}$) beams with
$\nu_{e}$($\bar{\nu}_{e}$), and the impossibility of separating CC
$\nu_{e}$($\bar{\nu}_{e}$) induced interactions from the NC
$\nu_{\mu}$($\bar{\nu}_{\mu}$) induced ones on an event by event
bases, the NuTeV result was obtained by performing a full
simulation of the whole experiment, in which the value of the weak
mixing angle was adjusted so that the Monte Carlo yield the best
description of the experimental data.  The NuTeV Monte Carlo
included, among may other things, a detailed simulation of the
neutrino(antineutrino) beam, a detailed model of the
$\nu(\bar{\nu})N$ cross section, QED radiative corrections,
charm-production-threshold effects, strange and charm sea
scattering, quasi-elastic scattering, neutrino-electron
scattering, non-isoscalar-target effects, higher twist effects,
etc \cite{zeller2}.  The experimental data and the Monte Carlo
results compare very well.

The NuTeV Monte Carlo simulation did not assume any asymmetry in
the strange-antistrange sea of the nucleons.
The effect of this asymmetry and possible effects due to nuclear medium
modifications to the parton distributions have been explored as possible
explanations for the discrepancy between the results for the
$\sin^{2}\theta_{W}$ obtained by NuTeV and those obtained by the global
fit to the DIS data \cite{LEP}.

The effect of the $s-\bar{s}\ $ asymmetry on the determination of
$\sin^{2}\theta_{W}$ can be accounted for through the use
of the Paschos-Wolfenstein (PW) relation \cite{PW}
\begin{equation}
R^{-} = \ \frac{\ \mbox{\large{$\sigma$}}_{NC}^{\nu N} -
\mbox{\large{$\sigma$}}_{NC}^{\bar{\nu}N}\ }{\
\mbox{\large{$\sigma$}}_{CC}^{\nu N} -
\mbox{\large{$\sigma$}}_{CC}^{\bar{\nu}N}\ } = \frac{1}{2} - \sin^{2}\theta_{W},
\label{PWR}
\end{equation}
where $\mbox{\large{$\sigma$}}_{N(C)C}^{\nu(\bar{\nu})N}$ refer to the
neutrino(antineutrino)-nucleon neutral(charged) current cross sections.

Equation~\ref{PWR} does not assume any asymmetry in the strange sea of the nucleon,
and has to be corrected when the target is a nucleus, due to effects of the
nuclear medium. In most experimental cases, as for
example the NuTeV experiment, the target is not isoscalar either and this effect
has to be taken into account.  A generalized PW relation
that includes all these modifications can easily be obtained. The extraction
of the $\sin^{2}\theta_{W}$ by NuTeV did not resort to any PW
relation because of the impossibility to separate effectively the
charged current form the neutral current signals.

Through the use of a generalized PW relation one can estimate the effect of
the presence of a nucleon-strange-sea asymmetry over
the extraction of the $\sin^{2}\theta_{W}$ done by NuTeV.
The same procedure can also allow to estimate the effect of different
nuclear medium modifications.

The generalized PW relation can be written as:
\begin{eqnarray}
\nonumber R^-&=&\frac{Z(\sigma_{NC}^{\nu p}-\sigma_{NC}^{\bar{\nu} p})+
N(\sigma_{NC}^{\nu n}-\sigma_{NC}^{\bar{\nu} n})}{Z(\sigma_{CC}^{\nu p}-
\sigma_{CC}^{\bar{\nu} p})+N(\sigma_{CC}^{\nu n}-\sigma_{CC}^{\bar{\nu} n})}\\
\nonumber &=& \frac{2}{(3N-Z)U^-+(3Z-N)D^-+3(N+Z)S^-}\times\\
\nonumber & & \left[\left(\frac{1}{4}-\frac{2}{3}\sin^{2}\theta_{W}\right)(ZU^-+ND^-)\right.\\
\nonumber  & & \left. +\left(\frac{1}{4}-\frac{1}{3}\sin^{2}\theta_{W}\right)(ZD^-+NU^-)\right.\\
\label{4-24} & & \left. +\left(\frac{1}{4}-\frac{1}{3}\sin^{2}\theta_{W}\right)(N+Z)S^-\right],
\end{eqnarray}
where,
\begin{eqnarray}
\label{4-12} U^-&=&\int^{1}_{0}x[u(x)-\bar{u}(x)]dx,\\
\label{4-13} D^-&=&\int^{1}_{0}x[d(x)-\bar{d}(x)]dx,\\
\label{4-14} S^-&=&\int^{1}_{0}x[s(x)-\bar{s}(x)]dx,
\end{eqnarray}
and $Z$ and $N$ are the proton and neutron numbers of the target nucleus.

The estructure functions measured in DIS experiments with nuclear targets
differ from those of the nucleon.  This is due to modifications caused by
the nuclear medium over the parton distributions.  The modified distributions
can be parametrized as \cite{Arneodo}
\begin{equation}
q^A_{i}(x,Q^2) = R^A_i(x,Q^2)q_i(x,Q^2).
\label{Nuclear}
\end{equation}
Parametrizations of the $R^A_i(x,Q^2)$ for different nuclei can be found
in the literature, for example those presented in reference~\cite{Eskola}
(available in CERNLIB).

In order to include nuclear-medium modifications in the generalized PW relation
it is enough to modify the asymmetry integrals as
\begin{eqnarray}
\label{4-33} U^{-(A)}&=&\int^{1}_{0}xR^A_u(x,Q^2)[u(x)-\bar{u}(x)]dx,\\
\label{4-34} D^{-(A)}&=&\int^{1}_{0}xR^A_d(x,Q^2)[d(x)-\bar{d}(x)]dx,\\
\label{4-35} S^{-(A)}&=&\int^{1}_{0}xR^A_s(x,Q^2)[s(x)-\bar{s}(x)]dx.
\end{eqnarray}

The NuTeV Collaboration extracted $\sin^{2}\theta_{W}$ from a full simulation
of the experiment in which a symmetric strange sea was assumed ($S^{-} = 0$).
From this, and the use of the generalized PW relation, one could determine
the value of $R^{-}$ consistent with the results of NuTeV, by
evaluating
\begin{align*}
 R^{-}_{\text{NuTeV}} &= R^{-}[S^-=0, Z, N,\sin^{2}\theta_{W}^{\text{NuTeV}} ]\\
 &= 0.2613,
\end{align*}
where N and Z correspond to the iron target used, and
$\sin^{2}\theta_{W}^{\text{NuTeV}}$ to the value reported by the
experiment.

Assuming $R^{-}_{\text{NuTeV}}$, the value of $\sin^{2}\theta_{W}$ from the global analysis 
of the DIS data ($\sin^{2}\theta_{W}^{\text{Global}}$), and the parametrizations of
Gluck, Reya and Vogt for the parton distributions of $u(x)$ and $d(x)$~\cite{GRV}, one 
can evaluate the level of asymmetry in the strange sea of the
nucleon that could explain the NuTeV result,
\begin{eqnarray*}
S^- &=& \frac{1}{\left[\frac{3}{2}R^{-}_{\text{NuTeV}}-\left(\frac{1}{4}-
\frac{1}{3}\sin^{2}\theta^{\text{Global}}_{W}\right)\right](N+Z)}\\
\nonumber & & \times \left[\left(\frac{1}{4}-\frac{2}{3}
\sin^{2}\theta^{\text{Global}}_{W}\right)(ZU^-+ND^-)\right.\\
\nonumber & & \left. + \left(\frac{1}{4}-\frac{1}{3}
\sin^{2}\theta^{\text{Global}}_{W}\right)(ZD^-+NU^-)\right.\\
\nonumber & & \left. -\frac{R^{-}_{\text{NuTeV}}}{2}[(3N-Z)U^-+(3Z-N)D^-]\right],\\
   &=& 0.004013.
\end{eqnarray*}

By including nuclear medium effects according to the parametrizations of 
Eskola \emph{et al.}~\cite{Eskola}, the value for the strange sea asymmetry that could 
explain the NuTeV result is
\begin{equation*}
S^{-{\text{(A)}}} = 0.003868.
\end{equation*}

The strange asymmetries predicted by our model, with and without
nuclear medium effects, are
\begin{eqnarray*}
S^{-}\ &=& 0.000087\,,\\
S^{-{\text{(A)}}} &=& 0.000047\,,
\end{eqnarray*}
which are two orders of magnitud smaller.  Since our parametrizations for
$x(s(x)-\bar{s(x)})$ are in agreement with the experimental data from Ref.
\cite{portheault}, one can conclude that the anomalous value for $\sin^{2}\theta_{W}$
reported by NuTeV cannot be explained in terms of a possible asymmetry in the strange
sea of the nucleon.

\section{Conclusions}
\label{conclu}

We have presented a model, based in fluctuations of the proton wavefunction to a generic 
Hyperon-Kaon Fock state, that closely reproduces experimental data on the extrange sea  
asymmetry of the nucleon. The model has a total of 8 parameters which have been fixed 
by fits to experimental data. No NNLO effects in the evolution of the $xs$ and $x\bar{s}$ 
have been considered, however the negative asymmetry introduced by NNLO evolution effects 
should be compensated by a large and positive asymmetry coming from the non-perturbative 
dynamics associated to the confining phase of QCD.

We investigated also the effect of such an asymmetry on the result presented by the 
NuTeV experiment on the measurement of $\sin^2 \theta_W$. In the study we considered, 
in addition, effects coming from the non isoscalarity of the NuTeV target and effects 
associated to the nuclear medium. Considering all together, we found that 
the effect of the $s-\bar{s}$ asymmetry in the nucleon sea is too small to account 
for the almost $3\sigma$ difference among the $\sin^2 \theta_W$ result by NuTeV and 
the world average. 

\section*{Acknowledgments}

J. Magnin would like to thanks the warm hospitality at the Physics Department, 
Universidad de los Andes, where part of this work was done. J.C. Sanabria would like to 
thanks also the warm hospitality during his visit to CBPF. This works was supported by 
the Brazilian Council for Science and Technology and FAPERJ (Brazil) under contract 
Project No.: E-26/170.158/2005.

\end{document}